\documentclass[aps,prb,%
twocolumn,
amsmath,amssymb,showpacs,
]{revtex4}

\usepackage[final]{graphicx}
\def\RCS$#1: #2 ${\expandafter\def\csname RCS#1\endcsname{#2}}
\RCS$Revision: 1.15 $ 
\RCS$Date: 2004/06/15 16:45:51 $ 
\RCS$Id: fxc-exciton.tex,v 1.15 2004/06/15 16:45:51 ralf Exp $

\newcommand{\tr}{\text{tr}}
\newcommand{\fxc}{f_{\text{xc}}}
\newcommand{\fxcQP}{f_{\text{xc}}^{\text{QP}}}
\newcommand{\fxcEx}{f_{\text{xc}}^{\text{Ex}}}
\newcommand{\vxc}{v_{\text{xc}}}
\newcommand{\Pixc}{\Pi_{\text{xc}}}

\newcommand{\GQP}{G_{\text{QP}}}

\newcommand{\chiS}{\chi_{\text{S}}}
\newcommand{\chiQP}{\chi_{\text{QP}}}

\newcommand{\VC}{V_{\text{C}}}

\newcommand{\rr}{\mathbf{r}}
\newcommand{\kk}{\mathbf{k}}
\newcommand{\qq}{\mathbf{q}}

\begin{document}

\title{Excitonic effects in time-dependent density-functional theory: An
analytically solvable model}

\author{R. Stubner}
\email[e-mail: ]{ralf.stubner@physik.uni-erlangen.de}

\author{I. V. Tokatly}
\author{O. Pankratov}

\affiliation{Lehrstuhl f\"ur Theoretische Festk\"orperphysik,
  Universit\"at Erlangen-N\"urnberg, Staudtstr.\ 7, 91058 Erlangen, Germany}

\date{15 June 2004}

\begin{abstract}
  We investigate the description of excitonic effects within
  time-dependent density-functional theory~(TDDFT). The
  exchange-correlation kernel $\fxc$ introduced in TDDFT
  allows a clear separation  of quasiparticle and
  excitonic effects. Using a diagrammatic representation for $\fxc$, we
  express its excitonic part $\fxcEx$ in terms of the effective vertex
  function~$\Lambda$. The latter fulfills an integral equation which
  thereby establishes the exact correspondence between TDDFT and the
  standard many-body approach based on Bethe-Salpeter equation~(BSE).The
  diagrammatic structure of the kernel in the equation for $\Lambda$
  suggests the possibility of strong cancellation effects. Should the
  cancellation take place, already the first-order approximation to
  $\fxcEx$ is sufficient. A potential advantage of TDDFT over the
  many-body BSE method is thus dependent on the efficiency of the
  above-quoted cancellation. We explicitly verify this for an
  analytically solvable two-dimensional two-band model. The calculations
  confirm that the low-order $\fxcEx$ perfectly describes the bound
  exciton as well as the excitonic effects in the continuous spectrum in
  a wide range of the electron--hole coupling strength.
\end{abstract}


\maketitle

\section{Introduction}
\label{sec:introduction}

Calculation of electronic excitation spectra remains one of the central
problems of the quantum theory of solids. Of special interest are
two-particle electron--hole excitations which determine the material's
optical properties. In semiconductors and insulators the electronic
screening is suppressed by the energy gap and the interaction of the
excited quasiparticles may substantially modify the excitation spectrum.
The excitonic effects stemming from this interaction comprise the
formation of bound electron--hole states as well as the alteration of
the absorption in the continuum spectrum above the band edge. The latter
is commonly referred to as unbound exciton effects or Sommerfeld
absorption enhancement.

In many-body perturbation theory two-particle excitations are
characterized by the two-particle Green function which satisfies the
Bethe-Salpeter equation~(BSE).\cite{Abr1963} Already in 1980 Hanke
and Sham\cite{Han1980}, using an approximate tight-binding
representation, showed that the BSE correctly describes the strong
excitonic features above the optical absorption edge in Si. In the
current ``state of the art'' procedure (see Ref.~\onlinecite{Oni2002}
for a recent review) the calculation of excitonic effects involves three
steps. First, a density-functional theory~(DFT) calculation in the
local density approximation~(LDA) is performed. On the second stage, the
LDA Kohn-Sham~(KS) energies and wavefunctions are used as a starting
point for the $GW$~calculation of the quasiparticle spectrum. Finally,
the BSE is solved numerically, using the $GW$ eigenvalues and the LDA
wavefunctions as input characteristics of the noninteracting
quasiparticles. The outlined procedure leads to highly accurate results,
as has been shown for a number of relatively simple systems, mostly bulk
semiconductors (see Ref.~\onlinecite{Oni2002} and references therein).
However, this method is extremely laborious, and for more complex
systems the calculations become prohibitively expensive.

A promising alternative, which is being intensively developed over
recent years, relies on time-dependent density-functional
theory~(TDDFT).\cite{Run1984} This theory allows to calculate (formally
exactly) the linear density--density response function and thereby the
excitation energies.\cite{Pet1996} Since in the framework of DFT the
exchange-correlation~(xc) effects are lumped in a local xc~potential
$\vxc$, the TDDFT equation for the response function contains the
variational derivative of $\vxc$ with respect to density
$\fxc(\rr,t;\rr',t') = \delta\vxc(\rr,t)/\delta n(\rr',t')$. This
xc~kernel $\fxc$ is the central unknown quantity of TDDFT in the linear
response regime. In their pioneering work Zangwill and
Soven\cite{Zan1980} calculated the photo-absorption in rare gases in a
self-consistent field manner. They used what later became known as
adiabatic local density approximation~(ALDA), simply substituting the
time-dependent density in the LDA xc~potential
$\vxc^{\text{ALDA}}=\vxc^{\text{LDA}}(n(\rr,t))$. The resulting
xc~kernel is local in space and time $\fxc^{\text{ALDA}} =
\delta(\rr-\rr') \delta(t-t') d\vxc^{\text{LDA}}/dn(\rr)$. ALDA has been
successfully applied to various finite systems like atoms or
molecules.\cite{Pet1996,Cas1998,Gis1998,Gra2000} 
 Typically in these systems already the RPA response
function calculated with KS eigenvalues and eigenfunctions gives good
results. The correction due to $\fxc^{\text{ALDA}}$ is quite small,
which signifies that Hartree effects dominate in the response function.
Unfortunately, $\fxc^{\text{ALDA}}$ remains insignificant also in
extended systems like semiconductors or insulators, where KS-RPA gives a
very poor description of the absorption spectra.\cite{Gav1997,Kim2003}
Thus whereas a correct accounting for xc~effects becomes crucial in
extended systems, the ALDA kernel $\fxc^{\text{ALDA}}$ fails to provide
even a reasonable starting approximation.

In the late nineties it has been realized that ALDA cannot serve as the
basis approximation for the dynamic xc~response of an inhomogeneous
electron gas, because of the intrinsically nonlocal nature of
$\fxc$.\cite{Vig1995b} For extended systems with an energy gap this
remains valid even in the static case.\cite{Gon1995,Gho1997}

The importance of the nonlocality of $\fxc$ was highlighted by the work
of Reining and coworkers\cite{Rei2002,Bot2004} who were able to describe
the contributions of unbound excitons in several diamond or zinc-blende
type semiconductors with a static xc~kernel proportional to
$1/|\rr-\rr'|$. Other examples for this are the exact exchange
kernel\cite{Kim2002a,Kim2002b} and the results of de Boeji \emph{et
  al.}\cite{Boe2001},  where in the context of
time-dependent current-density functional theory\cite{Vig1996,Vig1998}
the nonlocal effects were crucial for accurately describing the effects
of unbound excitons.\cite{Koo2000b} 

It can be easily understood that a nonlocal $\fxc$ is crucial
for describing excitonic effects. Within TDDFT the proper
polarization operator is defined via the RPA-like equation 
\begin{equation*}
  \tilde{\chi}(\omega) = \chiS(\omega) + \chiS(\omega) \cdot
                             \fxc(\omega) \cdot \tilde{\chi}(\omega)
  \text{,}
\end{equation*}
where in a crystalline solid $\tilde{\chi}$, the Kohn-Sham response
function $\chiS$, and $\fxc$ are matrices in reciprocal space. The
matrix structure of $\tilde{\chi}$ is responsible for local-filed
effects. However, these are relatively small in typical semiconductors
and can be neglected for a qualitative analysis. Keeping only diagonal
elements with zero reciprocal lattice vectors, we can easily solve for
$\tilde{\chi}$ obtaining 
\begin{displaymath}
  \tilde{\chi}(\omega,\qq) = \frac{\chiS(\omega,\qq)}%
                                  {1-\fxc(\omega,\qq)\chiS(\omega,\qq)}
  \text{.}
\end{displaymath}
The macroscopic dielectric function $\varepsilon_{\text{M}}$ is given by 
\begin{displaymath}
  \begin{split}
    \varepsilon_{\text{M}}(\omega) 
      &= 1 - \lim_{q \to 0} \VC(\qq) \tilde{\chi}(\omega,\qq) \\
      &= 1 - \lim_{q \to 0} \frac{\VC(\qq) \chiS(\omega,\qq)}%
                                 {1-\fxc(\omega,\qq)\chiS(\omega,\qq)} 
     \text{,}
\end{split}
\end{displaymath}
with the Coulomb interaction $\VC(\qq) = 4 \pi e^{2}/q^{2}$. An
additional excitonic peak in $\varepsilon_{\text{M}}(\omega)$ appears
when the denominator vanishes. However, it is well known that $\chiS$ is
proportional to $q^{2}$ in the limit $q \to 0$ for systems with an
energy gap.\cite{Adl1962,Wis1963} Hence $\fxc$ must behave as $1/q^{2}$
in this limit to counterbalance $\chiS$. Otherwise the xc~kernel would
have no effect on $\varepsilon_{\text{M}}(\omega)$ at all. For the
static long-ranged xc~kernel of Reining \emph{et al.} we have
$\fxc(\omega,\qq) = 4 \pi e^{2} \beta/q^{2}$ with some constant $\beta$.
The macroscopic dielectric function thus reads
\begin{displaymath}
  \varepsilon_{\text{M}} = 
    1 - \frac{4\pi e^{2} \alpha_{\text{S}}(\omega)}%
             {1 - 4\pi e^{2} \beta\alpha_{\text{S}}(\omega)}
  \text{,}
\end{displaymath}
where $\alpha_{\text{S}}(\omega)$ is the macroscopic polarizability of
the Kohn-Sham system $\alpha_{\text{S}}(\omega) = \lim_{q \to 0}
\chiS(\omega,\qq)/q^{2}$. For a typical $\alpha_{\text{S}}(\omega)$
close to the band edge this formula suggests the existence of only
\emph{one} excitonic peak. However, one expects several peaks from
unbound excitons above the band gap and bound excitons within
the gap. Phenomenologically one could overcome this problem by
introducing a frequency-dependent $\beta$. One though would need to
introduce very rapid oscillations in the region of the Rydberg series of
bound excitonic states .

Probably the most promising path in the quest for a good
approximation to $\fxc$ is a direct comparison of the TDDFT formalism
with the BSE.\cite{Rei2002,Sot2003,Adr2003,Mar2003} Simply comparing the
calculated spectra, it was found that it is often sufficient to use an
approximation to $\fxc$ which is of the first-order in the screened
particle--hole interaction. Although these results are very encouraging,
it is unclear why this approximation is so efficient and what its range
of validity is.

In this paper we derive a diagrammatic expression which exactly relates
the excitonic part of $\fxc$ to the BSE. We start with splitting
$\fxc$ into two parts separately accounting for quasiparticle and
excitonic effects. We then apply the diagrammatic rules we previously
derived\cite{Tok2002} to these two parts of $\fxc$. This leads us 
to an expression for the excitonic part of $\fxc$ in terms of the
three-point function $\Lambda$. The latter satisfies an integral
equation similar to the BSE which establishes the exact correspondence
between TDDFT and common many-body theory. The main advantage of this
approach is that the possibility of cancellation effects, which have been
conjectured in Ref.~\onlinecite{Mar2003}, is directly seen in the kernel of
the equation for $\Lambda$. 

In order to investigate the properties of our integral equation and the
applicability of low-order approximations we study a model two-band
system. In this model both the BSE and TDDFT equation can be solved
analytically, which offers an ideal test bed for approximations to the
exact $\fxc$. We find that indeed there are strong cancellation effects
in the integral equation for $\Lambda$ in the energy range close to the
band gap. For this reason both the shallow excitons and the unbound
excitonic effects are well described with a first-order
approximation to the excitonic part of $\fxc$.

The paper is organized as follows. In Sec.~\ref{sec:general} we
investigate diagrammatic properties of the excitonic part of $\fxc$ and
derive an exact correspondence between TDDFT and BSE. In
Sec.~\ref{sec:model} we introduce the model system that is used in
further calculations. Sections~\ref{sec:short-range-inter} and
\ref{sec:coulomb-interaction} are devoted to analytic calculations of
excitonic effects with a general short-ranged and the Coulomb
interaction respectively. Finally, we present our conclusions in
Sec.~\ref{sec:conclusion}.

\section{Diagrammatic meaning of $\fxc$ 
}
\label{sec:general}

\begin{figure}
  \includegraphics[width=0.8\linewidth]{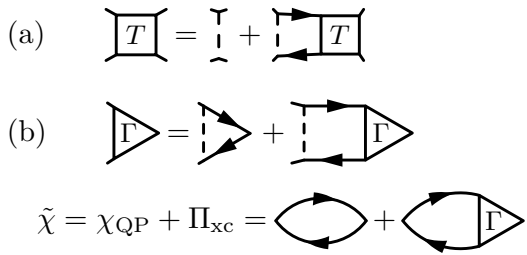}
  \caption{Diagrammatic representation of the Bethe-Salpeter equation}
  \label{fig:bse}
\end{figure}

The Bethe-Salpeter equation in the particle--hole channel is commonly
formulated as an integral equation for the particle--hole propagator or
the scattering~matrix $T$ in the ladder approximation. This integral
equation is equivalent to a summation of all ladder diagrams. The
diagrammatic representation of the $T$-matrix formulation is depicted in
Fig.~\ref{fig:bse}(a), where the full lines are the quasiparticle Green
functions and the dashed lines represent the screened interaction. In
the context of a connection to TDDFT we are not interested in the
four-point particle--hole propagator but rather in the response
functions. Therefore we consider a modified BSE where two of the
external lines of the scattering~matrix have been contracted to form the
three-point function $\Gamma$. The diagrammatic representation of the
integral equation for $\Gamma$ and its relation to the proper
polarization operator $\tilde{\chi}$ are depicted in
Fig.~\ref{fig:bse}(b). For our purposes this equation is equivalent to
the BSE. In the following we refer to the equation in
Fig.~\ref{fig:bse}(b) as the Bethe-Salpeter equation.

In terms of TDDFT the proper polarization operator $\tilde{\chi}$ is
given as 
\begin{equation}
  \label{eq:irred-polarization-tddft}
  \tilde{\chi}(\omega) = \chiS(\omega) + \chiS(\omega) \cdot
                             \fxc(\omega) \cdot \tilde{\chi}(\omega)
  \text{,}
\end{equation}
where $\chiS$ represents the density--density response function of the
noninteracting KS~particles, \emph{i.e.}, a bare loop of two
KS Green functions. Equation~\eqref{eq:irred-polarization-tddft} looks
like the RPA~equation, although it relates $\chiS$ with the full
$\tilde{\chi}$ which includes all self-energy and ladder diagrams. We
can therefore interpret $\fxc$ as an effective interaction which
describes self-energy and ladder diagrams in the annihilation channel.
Thereby $\fxc$ contains both quasiparticle and excitonic effects. As in
this paper we are only interested in the excitonic effects, it is
tempting to separate these two contributions to $\fxc$, as suggested in
previous works.\cite{Rei2002,Sot2003} This separation is indeed possible
without approximations, because we have
\begin{equation}
  \label{eq:fxc-splitting}
  \fxc = \chiS^{-1} - \tilde{\chi}^{-1}  
       = \underbrace{\chiS^{-1} - \chiQP^{-1}}%
                    _{=: \displaystyle \fxcQP} 
       + \underbrace{\chiQP^{-1} - \tilde{\chi}^{-1}}%
                    _{=: \displaystyle \fxcEx}
  \text{,}
\end{equation}
where $\chiQP$ is the density--density response function for
the noninteracting quasiparticles. By definition, $\fxcQP$ and
$\fxcEx$  are the kernels of the following RPA-type
equations:
\begin{subequations}
  \label{eq:dyson-fxcQP+fxcEX}
  \begin{align}
    \label{eq:dyson-fxcQP}
  \chiQP(\omega) &= \chiS(\omega) + \chiS(\omega) \cdot
                             \fxcQP(\omega) \cdot \chiQP(\omega) \\
    \label{eq:dyson-fxcEx}
  \tilde{\chi}(\omega) &= \chiQP(\omega) + \chiQP(\omega) \cdot
                             \fxcEx(\omega) \cdot \tilde{\chi}(\omega)
  \text{.}
  \end{align}
\end{subequations}
The newly introduced quantities $\fxcQP$ and $\fxcEx$ describe
quasiparticle and excitonic effects respectively. This can be visualized
by applying the diagrammatic rules for $\fxc$ as derived in
Ref.~\onlinecite{Tok2002}. The structure of the diagrammatic
representation of $\fxcQP$ and $\fxcEx$ is similar to the one for
$\fxc$, except that for $\fxcQP$ one has to use KS Green functions and
should only account for all possible self-energy insertions in every
order of the perturbation theory. This clearly describes the
quasiparticle corrections. For 
$\fxcEx$ one should use the quasiparticle Green functions with all
possible particle--hole interactions. Obviously, this reflects excitonic
contributions. It is important to note, that the general properties of
the perturbative expansion of $\fxc$ obtained in
Ref.~\onlinecite{Tok2002} remain valid separately for $\fxcQP$ and
$\fxcEx$. In particular, $\fxcQP$ remains finite at KS
excitation energies in every order of the perturbation theory. The same
holds for $\fxcEx$ at excitation energies of the noninteracting
quasiparticle system. Note that our $\fxcEx$ is the same as the
$\fxc^{\text{FQP}}$ of Refs.~\onlinecite{Sol2003} and \onlinecite{Adr2003}.

\begin{figure}
  \includegraphics[width=\linewidth]{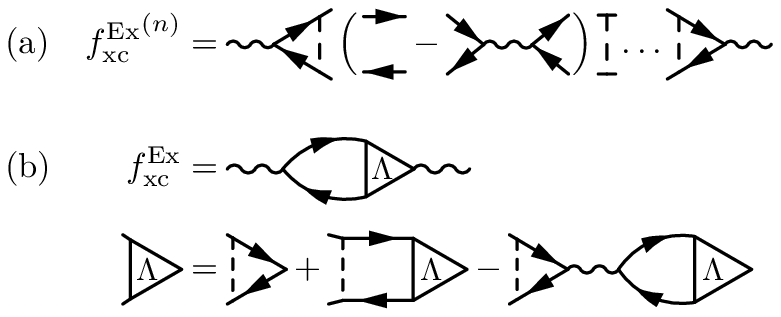}
  \caption{(a) The $n$-th order of $\fxcEx$ and (b)the integral equation
    for the three-point function $\Lambda$ which is an essential part of
    $\fxcEx$} 
  \label{fig:fxc-ex}
\end{figure}

Let us briefly outline the diagrammatic rules for the excitonic part of
$\fxc$. According to Ref.~\onlinecite{Tok2002} with the above-mentioned
modifications we have to draw all loops with $n$ particle--hole
interactions to construct the $n$-th order correction. These diagrams
serve as parent graphs for the construction of the $n$-th order $\fxcEx$.
To comply with the BSE we must use the ladder approximation here as
well.%
\footnote{In systems where the ladder approximation is not applicable
  one would have to replace the screened interaction (here and in the BSE)
  with the scattering amplitude.} 
Therefore, only one diagram with $n$ interactions is left. To the two
ends of the diagram we have to attach wiggled lines representing
$\chiQP^{-1}$. Next, we work out all possibilities to separate this
parent graph into two by cutting two fermionic lines. Then we join the
external fermionic lines of these parts, connect them by a wiggled line
and change the sign of the resulting diagram. Obviously, the only way to
separate the parent graph is to cut between adjacent interaction lines.
The cutting does not change the ladder structure of the diagrams as seen
in Fig.~\ref{fig:fxc-ex}(a). The summation of all ladder diagrams can be
cast in an integral equation as displayed in Fig.~\ref{fig:fxc-ex}(b).
If one inserts $\fxcEx$ obtained by solving the equation of
Fig.~\ref{fig:fxc-ex}(b) into Eq.~\eqref{eq:dyson-fxcEx} and calculates
the response function $\tilde{\chi}$, the result will be the same as the
$\tilde{\chi}$ obtained from the BSE. In this sense the equation of
Fig.~\ref{fig:fxc-ex}(b) gives the exact ``translation'' of the BSE into
the TDDFT language.

At this point we would like to highlight the differences between the
integral equation of Fig.~\ref{fig:fxc-ex}(b) and the equation for the
excitonic part of $\fxc$ derived in Ref.~\onlinecite{Mar2003} (Eq.~(4)
and (5) therein). The iterative equation of Marini \emph{et al.} is for
the two-point xc~kernel and is based on (finite order approximations to)
the xc~part of the response function and is logically analogous to our
diagrammatic expansion of $\fxc$ obtained in Ref.~\onlinecite{Tok2002}.
However, the equation of Fig.~\ref{fig:fxc-ex}(b) is an integral
equation for a three-point function analogous to the BSE%
\footnote{As noted above the BSE is normally formulated in terms of the
  four-point function $T$. Our integral equation for $\Lambda$ can
  straightforwardly be rewritten in terms of four-point functions as well.
  However, in the context of the present work it is more convenient to
  use three-point functions.}
and can be solved \emph{instead} of the BSE to obtain the same results. 

\begin{figure}
  \includegraphics[width=0.6\linewidth]{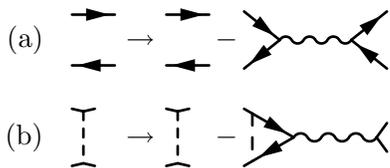}
  \caption{Diagrammatic representation of the two possible
    interpretations of the difference between BSE and our
    equation for $\Lambda$.}
  \label{fig:bse-lambda}
\end{figure}

Unfortunately, the exact calculation of $\fxcEx$ from the equation of
Fig.~\ref{fig:fxc-ex}(b) is at least as difficult as obtaining an exact
solution of the BSE. However, one can hope that the 
two-point kernel $\fxcEx$ is more suitable for approximations. An
indication in this direction can be seen directly from the diagrammatic
equation of Fig.~\ref{fig:fxc-ex}(b). Comparing this equation to
Fig.~\ref{fig:bse}(b) we see that it can be obtained from the BSE by
substitution of the particle--hole propagator as shown in
Fig.~\ref{fig:bse-lambda}(a). This is the same replacement that
was used in Ref.~\onlinecite{Tok2002} to prove the cancellation of
divergencies in $\fxc$ at KS excitation energies. Similarly, it
facilitates the cancellation of divergencies in $\fxcEx$ at QP
excitation energies.

Alternatively we can interpret the difference between the BSE and our
equation for $\Lambda$ as a modification of the interaction process in
the second diagram of the r.\,h.\,s.\ of the equation in
Fig.~\ref{fig:bse}(b). This change of the interaction is displayed in
Fig.~\ref{fig:bse-lambda}(b). As $\fxcEx$ is responsible for translating
the ladder diagrams into the annihilation channel, it is nontrivial only
if the ladder channel and the annihilation channel are distinguishable.
This is easily seen from the replacement of Fig.~\ref{fig:bse-lambda}(b). 
Indeed the ladder and the annihilation channel coincide for
nonrelativistic systems with a static point interaction. In this case
the quasiparticle propagators in Fig.~\ref{fig:bse-lambda}(b) form a
polarization loop which cancels the wiggled line. As a result the two
diagrams of Fig.~\ref{fig:bse-lambda}(b) cancel exactly. In the equation
for $\Lambda$ in Fig.~\ref{fig:fxc-ex}(b) this means that the last two
terms cancel and $\Lambda$ reduces to the first term. The excitonic part of
the xc~kernel then reduces to the interaction itself. One can
therefore expect that in systems with a short-ranged and almost static
effective interaction, the two terms of Fig.~\ref{fig:bse-lambda}(b)
will cancel to a large extent, and a low-order approximation to $\fxcEx$
will be sufficient. Conditions like this can, \emph{e.g.}, be found in
simple metals. On the contrary,
in semiconductors screening is less efficient and the effective
interaction is long-ranged. Further research is needed
to verify to what extent the cancellation is efficient.

Cancellation effects as we are expecting them from our diagrammatic
equation have been seen in Ref.~\onlinecite{Mar2003}. the success of the
lowest-order $\fxc$ found in Ref.~\onlinecite{Mar2003} implies that
\emph{de facto} the cancellation can be efficient in materials with a
band gap as well.

\section{Model System}
\label{sec:model}

In this section we consider a model system which reveals both bound and
unbound excitonic effects and where both the BSE and the
equation for $\Lambda$ can be solved analytically. With the exact
$\fxcEx$ at hand we can verify under what circumstances the
first-order approximation to $\fxcEx$ may be sufficient. The approximate
kernel must describe bound as well as unbound excitons. A simple
system with a bound exciton is given by the two-band Dirac model with a
static density--density interaction. Moreover, we consider
the two dimensional case in order to avoid 
technical difficulties with diverging integrals. The model Hamiltonian
is given by
\begin{multline}
  \label{eq:37}
  \mathcal{H} = \int d^{2}r \psi^{\dagger}(\rr) \hat{H} \psi(\rr) \\ 
              + \frac{1}{2} \int d^{2}r \int d^{2}r' 
                            \hat{n}(\rr)V(\rr-\rr')\hat{n}(\rr'),
\end{multline}
where $\hat{n}(\rr)=\psi^{\dagger}(\rr)\psi(\rr)$ is the density
operator, $\psi(\rr)$ is a two-component field operator, and $V(\rr-\rr')$
describes the interaction between the particles. The Hamiltonian for the
noninteracting particles reads
\begin{equation}
  \label{eq:model-hamiltonian}
  \hat{H} = \hat{k}_{x} \sigma_{x} + \hat{k}_{y} \sigma_{y} + \Delta \sigma_{z} =
  \left(%
    \begin{array}{cc}
      \Delta & \hat{k}_{-}  \\
      \hat{k}_{+}  & -\Delta \\
    \end{array}
  \right)
  \text{,}
\end{equation}
where $\sigma_{x,y,z}$ are the Pauli matrices,
$\hat{k}_{\pm} = \hat{k}_{x} \pm i \hat{k}_{y}$, the 2D momentum
operator  $\hat{\kk} = (\hat{k}_{x},\hat{k}_{y})$, and the band gap
equals to $2\Delta$. The 
energy dispersion for the noninteracting particles has two branches
which we label c and v for the (unoccupied) conduction band and the
(occupied) valence band:
\begin{equation}
  \label{eq:dispersion}
  E_{\text{c/v}}(\kk) = \pm E_{k} = \pm \sqrt{\Delta^{2} + k^{2}}
  \text{.}
\end{equation}
The eigenvectors of $\hat{H}$ are
\begin{equation}
  \label{eq:eigenvecs}
  \Psi_{\text{c}\kk} = \left(%
    \begin{array}{c}
      u_{k} \\ \frac{k_{+}}{k} v_{k}
    \end{array}
\right)
\text{,}\quad
  \Psi_{\text{v}\kk} = \left(%
    \begin{array}{c}
      - \frac{k_{-}}{k} v_{k} \\ u_{k}
    \end{array}
\right)
\end{equation}
with 
\begin{equation}
  \label{eq:uk-vk}
  u_{k} = \sqrt{\frac{1}{2}\left( 1 + \frac{\Delta}{E_{k}} \right)}
  \text{,}\quad
  v_{k} = \sqrt{\frac{1}{2}\left( 1 - \frac{\Delta}{E_{k}} \right)}
  \text{.}
\end{equation}

Note that this model can be understood as a two-component
(``relativistic'') system. The interactions in the ladder channel and in
the annihilation channel in such a system are always distinct regardless
whether the interaction is long- or short-ranged.  Therefore in this
model $\fxcEx$ is nontrivial even in the case of a point
interaction.

\begin{figure}
  \includegraphics[width=\linewidth]{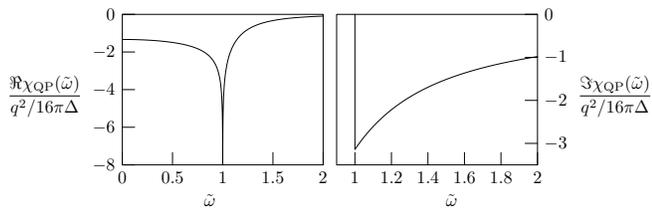}
  \caption{The real and the imaginary part of the exact $\chiQP$.}
  \label{fig:chiQP}
\end{figure}

We are going to solve this model in the ladder approximation,
\emph{i.e.}, ignoring self-energy terms and higher-order corrections to
the irreducible scattering matrix. Comparing to the BSE this implies
that $\hat{H}$ refers to the independent quasiparticles and $V$ is the
screened interaction between them. Therefore the one-particle Green
functions of $\hat{H}$ are the quasiparticle Green functions:
\begin{equation}
  \label{eq:GQP}
  \GQP(\omega,\kk) =
      \frac{\Psi_{\text{c}\kk}\Psi_{\text{c}\kk}^{\dagger}}%
            {\omega - E_{k} +i \delta} +
      \frac{\Psi_{\text{v}\kk}\Psi_{\text{v}\kk}^{\dagger}}%
            {\omega + E_{k} -i \delta}
  \text{.}
\end{equation}
Note that $\Psi_{\text{v}\kk}$ and $\Psi_{\text{c}\kk}$ are
two-component vectors and $\GQP$ is a $2\times2$~matrix. To lowest order
in the wave vector $\qq$ the quasiparticle response function 
$\chiQP(\omega,\qq)$ is given by
\begin{align}
  \label{eq:chiQP}
  \chiQP(\tilde{\omega},\qq) &= 
         -i\int \frac{d\epsilon}{2\pi} \int \frac{d^{2}k}{(2\pi)^{2}} 
           \tr \GQP(\epsilon+\tilde{\omega}, \kk+\qq) \GQP(\epsilon,\kk)
           \nonumber \\
  &= - \frac{q^{2}}{16\pi\Delta} \left(
    \frac{\tilde{\omega}^{2}+1}{2\tilde{\omega}^{3}}
    \ln\frac{1+\tilde{\omega}}{1-\tilde{\omega}} -
    \frac{1}{\tilde{\omega}^{2}}\right) 
  \text{,}
\end{align}
%
where $\tilde{\omega} = \omega/(2\Delta)$. The real and the imaginary
part of this function are displayed in Fig.~\ref{fig:chiQP}. A nonvanishing
imaginary part occurs only at frequencies above the quasiparticle gap
$\tilde{\omega}>1$, when the argument of the logarithm becomes
negative. Note that one sees here explicitly the $q^{2}$-dependence of
the response functions mentioned in the introduction.

In principle, we are now in the position to solve the BSE in this model
analytically.  However, to proceed further we need to introduce some technical
issues.  Due to the matrix structure of $\GQP$ one has to compute
traces when calculating $\chiQP$. Similarly, the three-point
functions~$\Gamma$ and $\Lambda$ are $2\times2$~matrices
and traces have to be calculated over internal indices in the diagrams
of Figs.~\ref{fig:bse}(a) and \ref{fig:fxc-ex}(b). Since all these matrices
have nonvanishing off-diagonal elements, evaluation of these traces
becomes quite tedious. Therefore it is convenient to choose the eigenstates of
$\hat{H}$ as the basis. The Green function $\GQP$ becomes then a diagonal
matrix with the elements 
\begin{equation}
  \label{eq:38}
  G_{\text{c/v}}(\omega,\kk) = \frac{1}{\omega \mp E_{k} \pm i\delta}
  \text{.}
\end{equation}
We thus can abandon the matrix notation altogether and use the two
scalar Green functions $G_{\text{c}}$ and $G_{\text{v}}$ instead of the
now diagonal $2 \times 2$ matrix $\GQP$. Now, every full
line is either a conduction- or a valence-state
propagator~\eqref{eq:38}. This, of course, increases the number of
diagrams we need to draw, because we have to consider all
possible combinations of conduction- and valence-band states. However,
in all diagrams the ``upper'' and ``lower'' Green functions that
constitute a particle--hole propagator must always be of a different
type (c or v). The diagrams with c--c or v--v two-particle propagators
vanish due to the integration over frequency since both
``upper'' and ``lower'' Green function have their pole in the same half of
the complex plane. This means that these diagrams do not contribute to
the polarization. The same holds for the three-point functions $\Gamma$
and $\Lambda$.  Instead of one equation for the $2\times2$~matrix
$\Gamma$ as depicted in Fig.~\ref{fig:bse}(b) we obtain two coupled
equations for the scalar functions $\Gamma_{\text{cv}}$ and
$\Gamma_{\text{vc}}$. For $\Gamma_{\text{cv}}$ the upper line is a
conduction-band state and the lower line a valence-band state, whereas
for $\Gamma_{\text{vc}}$ it is vice versa.

\begin{figure}
  \includegraphics[width=\linewidth]{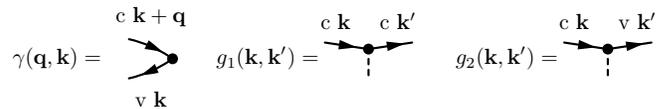}
  \caption{The diagrammatic representation of the ``bare'' vertex and
    the ``charges'' of Eqs.~\eqref{eq:bare-vertex}
    and~\eqref{eq:charges}.}  
  \label{fig:gamma-g}
\end{figure}

The transfer to the diagonal representation of $\GQP$ is equivalent to a
corresponding transformation of the field operators in
Eq.~\eqref{eq:37}. This transformation generates the ``bare'' vertex,
describing interaction with an external field in the polarization
diagrams, as well as the ``interaction vertices''. In the original
representation~\eqref{eq:model-hamiltonian} all these vertices are
simply unit matrices. They become, however, nontrivial matrices in the
diagonal representation. In fact, only three matrix elements of these
vertices are essential. These are depicted in Fig.~\ref{fig:gamma-g}. To
the lowest order in the transfered wave vector $\qq$ the bare vertex
computes to
\begin{equation}
  \label{eq:bare-vertex}
  \gamma(\qq,\kk) = \Psi_{\text{c} \kk+\qq}^{\dagger}\Psi_{\text{v} \kk} 
                  = \frac{1}{2 E_{k}} \left(
                          u_{k}^{2} q_{-} - 
                          \Bigl(v_{k} \frac{k_{-}}{k}\Bigr)^{2} 
                          q_{+} \right)
  \text{.}
\end{equation}
Note that $\gamma(\qq,\kk)$ is linear in $q$ because of the opposite
parity of c- and v-eigenfunctions at $\kk =0$. Thus the linear
$q$-dependence stems from the off-diagonal momentum
operator in the Hamiltonian~\eqref{eq:model-hamiltonian}. This is
actually the cause of the $q^{2}$-dependence of the response function we
refered to above in conjuction with Eq.~\eqref{eq:chiQP} and in the
introduction.  
The ``charges'', \emph{i.e.}, the vertices associated with the
interaction, are given by
\begin{subequations}
  \label{eq:charges}
  \begin{align}
    g_{1}(\kk,\kk') &= \Psi_{\text{c} \kk}^{\dagger}\Psi_{\text{c} \kk'}
                    = u_{k} u_{k'} + \frac{k_{-}}{k} v_{k} v_{k'} \frac{k'_{+}}{k'} \\
    g_{2}(\kk,\kk') &= \Psi_{\text{c} \kk}^{\dagger}\Psi_{\text{v} \kk'}
                    = \frac{k_{-}}{k} v_{k} u_{k'} - u_{k} v_{k'} \frac{k'_{-}}{k'}
   \text{.}
  \end{align}
\end{subequations}
All other possible combinations of valence- and
conduction-band states differ from Eqs~\ref{eq:bare-vertex} and
\ref{eq:charges} only by sign changes or complex conjugation.

From now on we use a diagram technique with two different types of full
lines representing conduction- and valence-band states. With these lines
we associate the scalar Green functions of Eq.~\eqref{eq:38}.  Vertices
are associated with the scalar functions of Eqs.~\eqref{eq:bare-vertex}
and~\eqref{eq:charges}.

There is an alternative interpretation of this basis transformation.
Consider one of the traces that has to be calculated for $\chiQP$
\begin{equation}
  \label{eq:34}
  \tr (\Psi_{\text{c}\kk+\qq}\Psi_{\text{c}\kk+\qq}^{\dagger})
      (\Psi_{\text{v}\kk}\Psi_{\text{v}\kk}^{\dagger})
  \text{,}
\end{equation}
where parenthesis show the grouping of the matrix multiplication. One
thus has to calculate the outer products of two vectors, multiply the
resulting matrices and in the end take the trace. However, this grouping
can be changed as follows
\begin{equation}
  \label{eq:35}
  \tr \Psi_{\text{c}\kk+\qq}(\Psi_{\text{c}\kk+\qq}^{\dagger}
      \Psi_{\text{v}\kk})\Psi_{\text{v}\kk}^{\dagger}
      =
  (\Psi_{\text{c}\kk+\qq}^{\dagger}\Psi_{\text{v}\kk})
  (\Psi_{\text{v}\kk}^{\dagger}\Psi_{\text{c}\kk+\qq})
  \text{.}
\end{equation}
Now one computes inner products of two vectors and multiplies the
resulting \emph{scalar} functions. Taking the trace is accounted
for automatically in Eq.~\eqref{eq:35}. The change of the diagram
technique outlined above is in effect a way to incorporate this
regrouping into the formalism.

As noted above, working in the basis of the conduction- and valence-band
states we have to split the BSE of Fig.~\ref{fig:bse}(b) into two scalar
equations for $\Gamma_{\text{cv}}(\omega,\qq,\kk)$ and
$\Gamma_{\text{vc}}(\omega,\qq,\kk)$. Note that while
$\Gamma_{\text{cv}}$ and $\Gamma_{\text{vc}}$ depend on two momenta,
they depend only on one frequency for a frequency-independent interaction.
These two three-point functions are in fact not independent but related
by the replacement $\qq \to -\qq$, $\omega \to -\omega$ and complex
conjugation. We can therefore derive one equation for
$\Gamma_{\text{cv}}$:
\begin{subequations}
  \label{eq:Gamma-cv}
\begin{align}
  \Gamma_{\text{cv}}(\omega,\qq,\kk) &=  \Gamma_{1}(\omega,\qq,\kk) \\
  &+
  \sum_{\kk'} V_{\kk,\kk'} 
  \frac{g_{1}(\kk,\kk')\Gamma_{\text{cv}}(\omega,\qq,\kk')g_{1}^{*}(\kk',\kk)}%
       {2 E_{k'} - \omega} \nonumber\\ 
  &+
  \sum_{\kk'} V_{\kk,\kk'} 
  \frac{g_{2}(\kk,\kk')\Gamma_{\text{cv}}^{*}(-\omega,-\qq,\kk')g_{2}(\kk',\kk)}%
       {2 E_{k'} + \omega} \nonumber \\
  \Gamma_{1}(\omega,\qq,\kk) &=  \sum_{\kk'} V_{\kk,\kk'} 
  \frac{g_{1}(\kk,\kk')\gamma(\qq,\kk')g_{1}^{*}(\kk',\kk)}%
       {2 E_{k'} - \omega} \\ 
  &+
  \sum_{\kk'} V_{\kk,\kk'} 
  \frac{g_{2}(\kk,\kk')\gamma^{*}(-\qq,\kk')g_{2}(\kk',\kk)}%
       {2 E_{k'} + \omega} \nonumber
  \text{.}
\end{align}
\end{subequations}
Note that we omit the $\qq$-dependence in the ``charges''
$g_{1}$ and $g_{2}$ as well as in the energy denominators, as we are
only interested in the lowest-order expansion in $\qq$, which stems from
the dipole matrix elements in the external vertices. Here $V_{\kk,\kk'}$
are the matrix elements of the interaction between the particles. Now we
investigate this equation for two different types of interaction, a
general short-range interaction and the long-ranged 
Coulomb interaction.

\section{A short-range interaction}
\label{sec:short-range-inter}

\subsection{The solution of the BSE}
\label{sec:solution}

In this section we solve Eq.~\eqref{eq:Gamma-cv} for a short-ranged
interaction, \emph{i.e.}, an interaction with a characteristic length
scale shorter than $\Delta^{-1}$. The final results are expressed in terms of the
physical (renormalized) scattering length, which includes the high
energy contribution to the integrals in Eq.~\eqref{eq:Gamma-cv}.  Having
in mind this renormalization we can formally use a momentum independent
bare interaction $V_{\kk,\kk'}=V$  in Eq.~\eqref{eq:Gamma-cv}.
Note that this does not make $\fxcEx$ trivially equal to the interaction
itself as discussed in the last paragraph of section~\ref{sec:general},
because for our two-band model the annihilation channel and the ladder
channel remain different even for a contact interaction. From
Eq.~\eqref{eq:Gamma-cv} we see, that the $\kk$-dependence of
$\Gamma_{\text{cv}}$ is given by the ``charges'' from
Eq.~\eqref{eq:charges} and has the same general form as for the bare
vertex \eqref{eq:bare-vertex}. The same is true for the 
$\qq$-dependence. Therefore we can write the following ansatz for
$\Gamma_{\text{cv}}$:
\begin{equation}
  \label{eq:ansatz}
  \Gamma_{\text{cv}}(\omega,\qq,\kk) = 
          u_{k}^{2} q_{-}
                    \Gamma_{\text{cv}}^{(s)}(\omega) + 
          \Bigl(v_{k} \frac{k_{-}}{k}\Bigr)^{2} q_{+}
                    \Gamma_{\text{cv}}^{(d)}(\omega)
  \text{.}
\end{equation}

Inserting this ansatz into Eq.~\eqref{eq:Gamma-cv} we obtain two coupled
equations: 
\begin{subequations}
  \label{eq:1}
  \begin{align}
    \Gamma_{\text{cv}}^{(s)}(\omega) =& V  \sum_{\kk} \frac{1}{2 E_{k}} 
         \left( \frac{u_{k}^{4}}{(2E_{k}-\omega)} - 
                \frac{v_{k}^{4}}{(2E_{k}+\omega)}\right) \nonumber\\ 
         &+ V \sum_{\kk} \frac{u_{k}^{4}}{2E_{k}-\omega}
                         \Gamma_{\text{cv}}^{(s)}(\omega) \nonumber\\
         &+ V \sum_{\kk} \frac{v_{k}^{4}}{2E_{k}+\omega} 
                         {\Gamma_{\text{cv}}^{(d)}}^{*}(-\omega)\\
    \Gamma_{\text{cv}}^{(d)}(\omega) =& V \sum_{\kk} \frac{1}{2 E_{k}} 
         \left( \frac{u_{k}^{4}}{(2E_{k}+\omega)} - 
                \frac{v_{k}^{4}}{(2E_{k}-\omega)}\right) \nonumber\\ 
         &+ V \sum_{\kk} \frac{v_{k}^{4}}{2E_{k}-\omega} 
                         \Gamma_{\text{cv}}^{(d)}(\omega)\nonumber\\
         &+ V \sum_{\kk} \frac{u_{k}^{4}}{2E_{k}+\omega}
                         {\Gamma_{\text{cv}}^{(s)}}^{*}(-\omega)
    \text{.} 
  \end{align}
\end{subequations}
From these equations it immediately follows that
$\Gamma_{\text{cv}}^{(s)}(\omega) =
{\Gamma_{\text{cv}}^{(d)}}^{*}(-\omega)$ and the equation to solve
reduces to
\begin{align}
  \label{eq:2}
    \Gamma_{\text{cv}}^{(s)}(\omega) =& V \sum_{\kk} \frac{1}{2 E_{k}} 
         \left( \frac{u_{k}^{4}}{(2E_{k}-\omega)} - 
                \frac{v_{k}^{4}}{(2E_{k}+\omega)}\right) \nonumber\\ 
         &+ V \sum_{\kk} \left(\frac{u_{k}^{4}}{2E_{k}-\omega} +
                               \frac{v_{k}^{4}}{2E_{k}+\omega}  \right)
                         \Gamma_{\text{cv}}^{(s)}(\omega)\nonumber\\
        =:& V \gamma_{1}(\omega) + V K_{0}(\omega) 
                                    \Gamma_{\text{cv}}^{(s)}(\omega)
    \text{.}
\end{align}

At this point it is convenient to perform the above-mentioned
renormalization of the interaction by splitting $K_{0}$ into a low- and
a high-energy part, the latter being equal to $\sum_{\kk}\frac{1}{4E_{k}}$. The high
energy part logarithmically diverges at large $k$. This divergence can
be removed by the standard renormalization of the interaction 
\begin{equation}
  \label{eq:3}
  \tilde{V} = \frac{V}{1-V\sum_{\kk}\frac{1}{4E_{k}}} =: \frac{4\pi}{\Delta}a
  \text{,}
\end{equation}
where we introduce the dimensionless scattering length $a$. With
this renormalized interaction, $\Gamma_{\text{cv}}^{(s)}(\omega)$
fulfills the following equation
\begin{equation}
  \label{eq:4}
  \Gamma_{\text{cv}}^{(s)}(\omega) = \tilde{V} \gamma_{1}(\omega) + 
                                     \tilde{V} \tilde{K}_{0}(\omega) 
                                          \Gamma_{\text{cv}}^{(s)}(\omega)
\end{equation}
with
\begin{equation}
  \label{eq:5}
  \tilde{K}_{0}(\omega) = K_{0}(\omega) -\sum_{\kk}\frac{1}{4E_{k}} 
                        = \sum_{\kk} \frac{(2\Delta+\omega)^{2}}%
                                          {4E_{k}(4E_{k}^{2}-\omega^{2})}
  \text{.}
\end{equation}
Equation~\eqref{eq:4} has the obvious solution
\begin{equation}
  \label{eq:6}
  \Gamma_{\text{cv}}^{(s)}(\omega) = \frac{\tilde{V}\gamma_{1}(\omega)}%
                                          {1-\tilde{V}\tilde{K}_{0}(\omega)}
  \text{.}
\end{equation}
Calculating the 2D integrals in Eq.~\eqref{eq:5} and
$\gamma_{1}(\omega)$ in Eq.~\eqref{eq:2} we obtain
\begin{equation}
  \label{eq:7}
  \tilde{K}_{0}(\tilde{\omega}) = \frac{\Delta}{4\pi}
                          \frac{(1+\tilde{\omega})^{2}}{4\tilde{\omega}}
                          \ln\frac{1+\tilde{\omega}}{1-\tilde{\omega}} 
                        =: \frac{\Delta}{4\pi} F(\tilde{\omega})
\end{equation}
and
\begin{align}
  \label{eq:8}
  \gamma_{1}(\tilde{\omega}) &= 
                       \frac{1}{16\pi}
                       \left(\frac{(1+\tilde{\omega})^{2}}%
                                  {2\tilde{\omega}^{2}}
                             \ln\frac{1+\tilde{\omega}}{1-\tilde{\omega}} 
                             -\frac{1}{\tilde{\omega}}\right) \nonumber\\
                    &= \frac{1}{16\pi} \frac{2}{\tilde{\omega}}
                       \left(F(\tilde{\omega})-\frac{1}{2}\right)
  \text{.}
\end{align}

\begin{figure}[tbp]
  \includegraphics[width=0.6\linewidth]{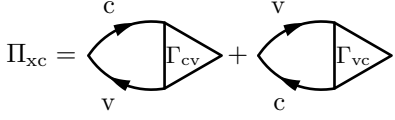}
  \caption{The diagrammatic expression for $\Pixc$.}
  \label{fig:pixc-diag}
\end{figure}

Inserting these integrals into the solution~\eqref{eq:4} we can compute
the response function's xc part $\Pixc=\tilde{\chi}-\chiQP$ and its
first-order approximation $\Pixc^{(1)}$. In diagrammatic form $\Pixc$ is
displayed in Fig.~\ref{fig:pixc-diag}. The results are
\begin{widetext}
\begin{equation}
  \label{eq:9}
  \Pixc(\tilde{\omega},\qq) = - \frac{q^{2}}{16\pi\Delta}
                              \frac{a}{\tilde{\omega}^{2}}
                              \left(
                              \frac{\left(F(\tilde{\omega})-\frac{1}{2}\right)^{2}}%
                                   {1-aF(\tilde{\omega})} +
                              \frac{\left(F^{*}(-\tilde{\omega})-\frac{1}{2}\right)^{2}}%
                                   {1-aF^{*}(-\tilde{\omega})}
                              \right)
\end{equation}
and
\begin{equation}
  \label{eq:10}
  \Pixc^{(1)}(\tilde{\omega},\qq) = - \frac{q^{2}}{16\pi\Delta}
                              \frac{a}{\tilde{\omega}^{2}}
                              \left( \textstyle
                              \left(F(\tilde{\omega})-\frac{1}{2}\right)^{2}+
                              \left(F^{*}(-\tilde{\omega})-\frac{1}{2}\right)^{2}
                              \right)
  \text{,}
\end{equation}
\end{widetext}
where the function $F(\omega)$ is defined in Eq.~\eqref{eq:7}.

The equation for $\Lambda$ given in Fig.~\ref{fig:fxc-ex}(b) can be
solved in a similar fashion, which allows then to calculate the excitonic
part of the exact xc kernel. However, it is easier to obtain $\fxcEx$
directly from Eq.~\eqref{eq:fxc-splitting} and the exact response
function $\tilde{\chi} = \chiQP + \Pixc$ with $\Pixc$ from
Eq.~\eqref{eq:9}: 
\begin{equation}
  \label{eq:fxc-exact}
  \fxcEx(\tilde{\omega},\qq) = \frac{\chiQP^{-1}(\tilde{\omega},\qq)
                                          \Pixc(\tilde{\omega},\qq) 
                                          \chiQP^{-1}(\tilde{\omega},\qq)}%
                                  {1 + \chiQP^{-1}(\tilde{\omega},\qq) 
                                          \Pixc(\tilde{\omega},\qq)}
  \text{.}
\end{equation}
From here the first-order approximation to $\fxcEx$ immediately follows: 
\begin{equation}
  \label{eq:fxc-first}
  {\fxcEx}^{(1)}(\tilde{\omega},\qq) = \chiQP^{-1}(\tilde{\omega},\qq) 
                                      \Pixc^{(1)}(\tilde{\omega},\qq) 
                                      \chiQP^{-1}(\tilde{\omega},\qq)
  \text{.}
\end{equation}
Inserting this ${\fxcEx}^{(1)}$ in Eq.~\eqref{eq:dyson-fxcEx}
we arrive at an approximate solution for the response function's xc part 
\begin{equation}
  \label{eq:pixc-1-approx}
  \Pixc^{f^{(1)}}(\tilde{\omega},\qq) =
                      \frac{\Pixc^{(1)}(\tilde{\omega},\qq)}%
                           {1-\Pixc^{(1)}(\tilde{\omega},\qq)
                              \chiQP^{-1}(\tilde{\omega},\qq)}
\end{equation}
Note that although ${\fxcEx}^{(1)}$ is based on $\Pixc^{(1)}$, this
formula does \emph{not} coincide with $\Pixc^{(1)}$. In accordance with 
Eq.~\eqref{eq:dyson-fxcEx} it accounts for an infinite series of
diagrams instead. This way, the excitonic pole in $\Pixc$,
Eq.~\eqref{eq:9}, which has been lost in $\Pixc^{(1)}$,
Eq.~\eqref{eq:10}, reappears in Eq.~\eqref{eq:pixc-1-approx}.

\subsection{Results}
\label{sec:results}

Having calculated the exact and the approximate expressions for the
excitonic part of the xc~kernel and for the response function we are now
in the position to compare these results. Let us start with the
excitonic peak in the absorption spectrum. In both the exact and the
approximate response functions the excitonic peak originates from the
divergence of the xc~part. The exact $\Pixc$ in Eq.~\eqref{eq:9} has a
pole at
\begin{equation}
  \label{eq:11}
  1-aF(\tilde{\omega}) = 0
\end{equation}
for positive $a$ and $0 \le \tilde{\omega} \le 1$.
When the dimensionless exciton binding energy $\tilde{\varepsilon} =
1-\tilde{\omega}$ is small, Eq.~\eqref{eq:11} has an approximate
solution 
\begin{equation}
  \label{eq:12}
  \tilde{\varepsilon} = 2 \exp\left( -\frac{1}{a} \right)
  \text{.}
\end{equation}
The approximate response function of Eq.~\eqref{eq:pixc-1-approx} which
is based on ${\fxcEx}^{(1)}$ has a pole 
\begin{equation}
  \label{eq:13}
  1-\chiQP^{-1}(\tilde{\omega},\qq) \Pixc^{(1)}(\tilde{\omega},\qq) = 0
  \text{.}
\end{equation}
Similarly to the solution~~\eqref{eq:12} of the exact
equation~\eqref{eq:11}, for a small binding energy $\tilde{\varepsilon}$
this equation can be approximately solved by
\begin{equation}
  \label{eq:14}
  \tilde{\varepsilon} = 2 \exp\left( -\frac{1}{a} + 
                        \frac{1-a-\sqrt{1-2a-a^{2}}}{2a} \right)
  \text{,}
\end{equation}
which differs from the exact result~\eqref{eq:12}. However, \emph{e.g.}
at $a = 0.2$ this is only about 14\,\% larger than the exact
solution~\eqref{eq:12}, which gives $\tilde{\varepsilon} \approx 0.013$.
For a smaller scattering length and therefore a smaller binding energy
the agreement between the exact result and the one based on ${\fxcEx}^{(1)}$
is even better.

\begin{figure}
  \includegraphics[width=\linewidth]{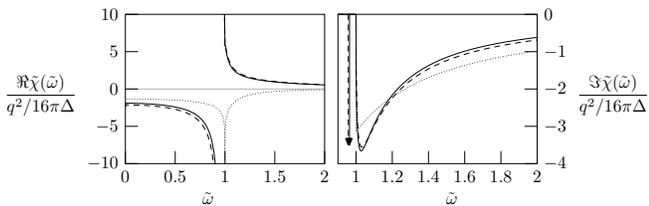}
  \caption{The real and the imaginary parts of the exact $\tilde{\chi}$
    (full) and the $\tilde{\chi}$ based on ${\fxcEx}^{(1)}$ (dashes) for
  the scattering length $a=0.25$. The real and the imaginary part of
  $\chiQP$ are shown by the dotted line.} 
  \label{fig:chi-0.25}
\end{figure}

In Fig.~\ref{fig:chi-0.25} the real and the imaginary part of the exact
$\tilde{\chi}$ and the $\tilde{\chi}$ based on ${\fxcEx}^{(1)}$ are
shown for the scattering length $a=0.25$. For comparison, $\chiQP$ is
also displayed. One clearly observes a very good agreement
between the exact and the approximate response functions. Both main
features, the enhancement of the imaginary part (the Sommerfeld factor) and
the excitonic peak are correctly reproduced. The latter is indicated by
the arrows on the plots of the imaginary part. The apparent difference
between the exact and the approximated response function is the exact
position of the excitonic peak as described above. The oscillator
strengths should also be different, which is not, however, reflected in
the figure.

\begin{figure}
  \includegraphics[width=\linewidth]{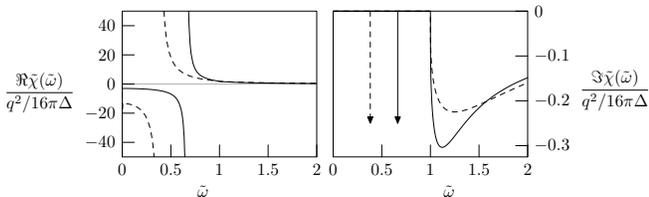}
  \caption{The real and the imaginary parts of the exact $\tilde{\chi}$
    (full) and the $\tilde{\chi}$ based on ${\fxcEx}^{(1)}$ (dashes) for
  the scattering length $a=0.6$} 
  \label{fig:chi-0.6}
\end{figure}

From the above discussion one could conjecture that the approximation
for the response function based on ${\fxcEx}^{(1)}$ is always sufficient.
This is, of course, not true, as we can see from
Fig.~\ref{fig:chi-0.6}, where the approximate and the exact
response function are compared for the larger value of the scattering
length $a=0.6$. With this large interaction%
\footnote{The natural scale for the interaction strength in our system
  is given by the value $a=2$. At this value for $a$ the exciton binding
  energy is equal to the band gap, which allows for a spontaneous
  creation of excitons (excitonic instability) in the system.} 
the position of the excitonic peak in the approximate response function
is clearly wrong. In addition, the imaginary part has the wrong
magnitude above the band gap.

\begin{figure}
  \includegraphics[width=\linewidth]{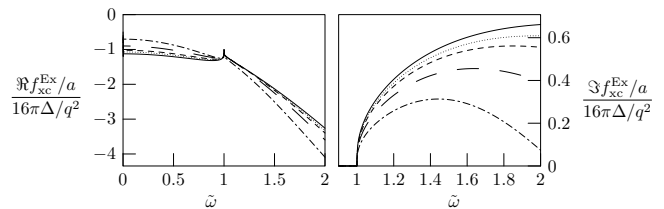}
  \caption{The real and the imaginary part of the exact $\fxcEx/a$ for
    $a=0.05$ (dots), $a=0.1$ (dashes), $a=0.25$ (long dashes), and
    $a=0.6$ (dot-dashes) compared with ${\fxcEx}^{(1)}/a$ (full).} 
  \label{fig:fxc-full}
\end{figure}

To uncover the background of the good agreement between the exact and
the approximate response functions for a ``weak'' interaction, let us
compare ${\fxcEx}^{(1)}$ with the exact $\fxcEx$ for different interaction
strengths. We note first that ${\fxcEx}^{(1)}$ is proportional to the
scattering length $a$. Therefore, more adequate is to compare
${\fxcEx}^{(1)}/a$ with $\fxcEx/a$, as done in Fig.~\ref{fig:fxc-full}.
It is clearly visible that ${\fxcEx}^{(1)}$ is a very good approximation
to the exact $\fxcEx$ in a frequency range close to the band gap.
Actually, for $\tilde{\omega} = 1$ we obtain from our exact expressions
\begin{equation}
  \label{eq:29}
  {\fxcEx}^{(1)}(\tilde{\omega} = 1) =
    - \frac{16\pi\Delta}{q^{2}} a 
\end{equation}
and
\begin{equation}
  \label{eq:30}
  \fxcEx(\tilde{\omega} = 1) = 
     - \frac{16\pi\Delta}{q^{2}} \frac{a}{1-a^{2}/4} 
  \text{.}
\end{equation}
Thus even for a very strong interaction of $a=0.6$ where
${\fxcEx}^{(1)}$ leads to the rather poor response function of 
Fig.~\ref{fig:chi-0.6}, there is only a 10\,\% error in
${\fxcEx}^{(1)}$ at $\tilde{\omega} = 1$. In the static case the errors
are larger, as we get for $\tilde{\omega}=0$:
\begin{equation}
  \label{eq:31}
  {\fxcEx}^{(1)}(\tilde{\omega}=0) = -\frac{16\pi\Delta}{q^{2}}
                                     \frac{9}{8} a
\end{equation}
and
\begin{equation}
  \label{eq:32}
  \fxcEx(\tilde{\omega}=0) = -\frac{16\pi\Delta}{q^{2}}
                             \frac{9}{8} \frac{a}{1+a}
\text{.}
\end{equation}
Here a 10\,\% error is already reached for $a=0.1$, as the ratio of
 $\fxcEx$ and ${\fxcEx}^{(1)}$ is of order $a$, whereas it is of order
 $a^{2}$ at $\tilde{\omega} = 1$. 

\begin{figure}
  \includegraphics[width=\linewidth]{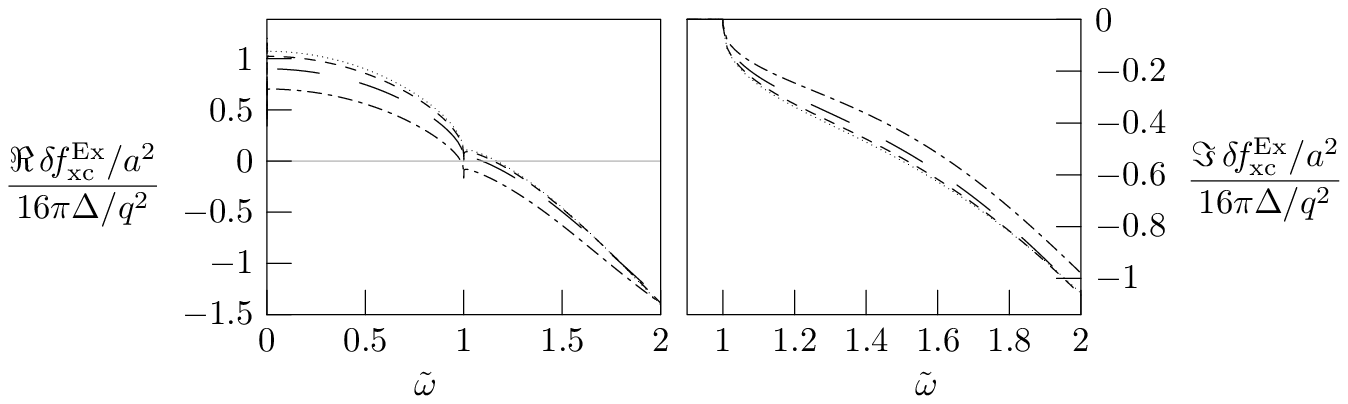}
  \caption{The real and the imaginary parts of $\delta\fxcEx/a^{2} =
    (\fxcEx - {\fxcEx}^{(1)})/a^{2}$ for $a=0.05$ (dots), $a=0.1$
    (dashes), $a=0.25$ (long dashes), and $a=0.6$ (dot-dashes).} 
  \label{fig:fxc-diff}
\end{figure}

It is also interesting to look at the difference $\delta\fxcEx$ between
the exact $\fxcEx$ and ${\fxcEx}^{(1)}$ for different interaction
strengths. Since the leading order term in $\delta\fxcEx$ is of the
order $a^{2}$, we normalize these differences by $a^{2}$, when plotting
them in Fig.~\ref{fig:fxc-diff}. The general behavior of the different
curves is quite similar, which indicates that it is mostly the second-order
approximation ${\fxcEx}^{(2)}$ which contributes to $\delta\fxcEx$. One
may be surprised that the curves for stronger interaction are closer to
$0$ for $\tilde{\omega}<1$, which might indicate that ${\fxcEx}^{(1)}$
is a better approximation for \emph{stronger} interactions. However,
this is not the case. This only tells us that ${\fxcEx}^{(3)}$ is
positive for $\tilde{\omega}<1$.

\subsection{The validity of the first-order approximation}
\label{sec:interpretation}

From the previous section we can conclude that ${\fxcEx}^{(1)}$ is a
good approximation to $\fxcEx$ close to the band gap $\tilde{\omega}=1$
practically for any interaction strength.  Even for a strong
interaction, where the response function stemming from ${\fxcEx}^{(1)}$
is quite wrong, the first-order kernel ${\fxcEx}^{(1)}$ is still very
good close to $\tilde{\omega}=1$. This is the reason for the success of
${\fxcEx}^{(1)}$ in describing the bound excitonic states as shown in
the previous section. Since around $\tilde{\omega}=1$ 
the first-order kernel ${\fxcEx}^{(1)}$ is a good
approximation, if the (exact) bound excitonic state lies
within this region, ${\fxcEx}^{(1)}$ will describe it correctly. If,
however, the binding energy is outside this region, ${\fxcEx}^{(1)}$
will fail. Note that as seen from Fig.~\ref{fig:fxc-full} this region
gets smaller as the interaction increases and at the same time the
binding energy of the exciton increases. Hence the error in the exciton
binding energy increases with the increase of the interaction strength. 

\begin{figure}[h]
  \includegraphics[width=\linewidth]{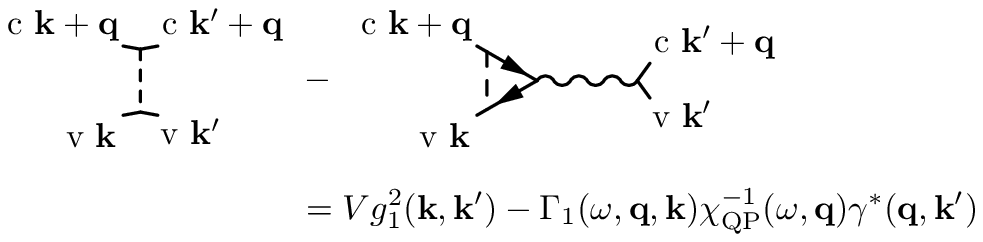}
  \caption{Explicit calculation of one replacement. Here $\omega$ and
    $\qq$ are the transfered energy and momentum.}
  \label{fig:explicit-replacement}
\end{figure}

Can we understand the good agreement between ${\fxcEx}^{(1)}$ and $\fxcEx$
for $\tilde{\omega} \approx 1$ in terms of the integral equation for
$\Lambda$? For this we explicitly calculate the diagrams
involved in the replacement shown in Fig.~\ref{fig:bse-lambda}(b).
Working with separate Green functions for the valence- and the
conduction-band states we have four possible combinations of 
the external lines and can therefore split this replacement into four
parts according to these combinations. At energies close to the band gap
the two-particle propagator with a conduction-band state in the upper line
should dominate. Hence, the most important part of this replacement is
the diagram with a conduction-band state in the upper line on both sides
of the graphs. This diagram together with its translation into
quantities introduced in the previous section is displayed in
Fig.~\ref{fig:explicit-replacement}. Note that we use the bare
interaction $V$ here, as the interaction renormalization in these
diagrams means simply a replacement $V \to \tilde{V}$.

Inserting $g_{1}$, $\Gamma_{1}$, $\chiQP$, and $\gamma$ in the
expression in Fig.~\ref{fig:explicit-replacement} and taking the limit
$\tilde{\omega}\to1$ we obtain: 
\begin{widetext}
\begin{multline}
  \label{eq:36}
  V g_{1}^{2}(\kk,\kk') - \Gamma_{1}(\tilde{\omega}=1,\qq,\kk)
           \chiQP^{-1}(\tilde{\omega}=1,\qq) \gamma^{*}(\qq,\kk') = \\
  V \biggl( u_{k}^{2} u_{k'}^{2} \Bigl(1-\frac{\Delta}{E_{k'}}\Bigr)
           + \Bigl(\frac{k_{-}}{k}v_{k}\Bigr)^{2}
             \Bigl(\frac{k'_{+}}{k'}v_{k'}\Bigr)^{2} 
           + 2 \frac{k_{-}}{k}v_{k}u_{k} u_{k'}v_{k'}\frac{k'_{+}}{k'}
           + u_{k}^{2} \Bigl(\frac{k'_{+}}{k'}v_{k'}\Bigr)^{2}
             \Bigl(\frac{q_{-}}{q}\Bigr)^{2} \frac{\Delta}{E_{k'}}
    \biggr)
\text{.}
\end{multline}
\end{widetext}
Not surprisingly the whole expression~\eqref{eq:36} is proportional to
the interaction $V$ (or $\tilde{V}$ after renormalization). In the
proportionality factor in round brackets all summands contain
$(1-\Delta/E_{k'})$ (or powers 
thereof). For the first summand this is directly visible, in the others
it is ``hidden'' in $v_{k'}$.  Let us look at this factor more closely.
When the diagram of Fig.~\ref{fig:explicit-replacement} is 
inserted into the equation for $\Lambda$ of Fig.~\ref{fig:fxc-ex}(b) the
integration over $\kk'$ has to be performed. The main contribution to
this integral comes from small momenta, due to the small energy denominators in
the particle--hole propagator. However, for these small momenta the
expression~\eqref{eq:36} is small as $\Delta \approx E_{k'}$.
 
We can therefore conclude that close to the band gap the kernel in the
equation for $\Lambda$ is indeed small for those states which 
mainly contribute to the integral. Thus we explicitly observe the
cancellation we qualitatively discussed in section~\ref{sec:general}.
This cancellation explains the excellent agreement between $\fxcEx$ and
${\fxcEx}^{(1)}$ for $\tilde{\omega} \approx 1$.

\begin{figure}
  \includegraphics[width=\linewidth]{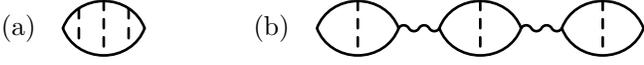}
  \caption{Third-order term in the expansion of (a) the exact
    $\tilde{\chi}$ and (b) the $\tilde{\chi}$ based on ${\fxcEx}^{(1)}$
    in terms of the interaction.}
  \label{fig:chi3-comparison}
\end{figure}

A complimentary interpretation can be obtained from looking
at the diagrammatic expansion of the response functions. In
Fig.~\ref{fig:chi3-comparison} the third-order term in the expansion of 
the exact $\tilde{\chi}$ and the $\tilde{\chi}$ based on
${\fxcEx}^{(1)}$ is displayed. The 
difference between the two expressions is similar to the replacements
discussed in section~\ref{sec:general}. This remains true in all orders
of the perturbation theory. From the previous calculation of the effect
of this replacement follows, that the exact $\tilde{\chi}$ and
the $\tilde{\chi}$ based on ${\fxcEx}^{(1)}$ are almost identical
in the vicinity of the band gap, independently of the position of the
excitonic peak. 

\section{The Coulomb interaction}
\label{sec:coulomb-interaction}

In this section, we consider the 2D~Dirac model of Eq.~\eqref{eq:37}
with the Coulomb interaction between the particles, \emph{i.e.},
\begin{equation}
  \label{eq:15}
  V(\rr-\rr') = \frac{e^{2}}{|\rr-\rr'|}
  \text{,}
\end{equation}
where $e$ is the particle's charge. Note that Eq.~\eqref{eq:15} is the
3D interaction although our model system is 2D. We will not attempt to
solve exactly the BSE with this interaction, but rather 
focus on the properties of shallow excitons. The weak binding limit
$\tilde{\varepsilon} = 1-\tilde{\omega} \ll 1$ is, in fact, a
``nonrelativistic'' limit where the BSE reduces
to the two-particle Schr\"odinger equation.\cite{Sha1966}
The quasiparticle energy eigenvalues are approximately
\begin{equation}
  \label{eq:16}
  E_{k} \approx \Delta + \frac{k^{2}}{2\Delta}
  \text{,}
\end{equation}
and for the eigenvectors holds
\begin{equation}
  \label{eq:17}
  u_{k} \approx 1 \quad\text{and}\quad v_{k} \approx 0
  \text{.}
\end{equation}
Solving the BSE thus reduces to solving the positronium problem in 2D
for particles with a mass $\Delta$, \emph{i.e.}, 
for a reduced mass $\Delta/2$. The response function can be written in
the spectral representation as\cite{Mahan1990} 
\begin{align}
  \label{eq:18}
  \tilde{\chi}(\omega,\qq) &= |\gamma(\qq,0)|^{2} \sum_{n} 
                                   \frac{|\psi_{n}(r=0)|^{2}}%
                                        {\omega-(2\Delta-\varepsilon_{n})} 
                                        \nonumber\\
                       &\approx |\gamma(\qq,0)|^{2}
                                   \frac{|\psi_{0}(r=0)|^{2}}%
                                        {\omega-(2\Delta-\varepsilon_{0})}
  \text{,}
\end{align}
where the $\psi_{n}$ are the eigenfunctions with eigenvalues
$\varepsilon_{n}$ (positive for bound states) of the above-mentioned Schr\"odinger
equation. The approximation in Eq.~\eqref{eq:18} is valid
close to the excitonic peak of the $1s$ ground state, in which we are 
interested in here. The $1s$ wavefunction needed in this approximation
is given by 
\begin{equation}
  \label{eq:19}
  \psi_{0}(r) =   \sqrt{\frac{2\Delta\varepsilon_{0}}{\pi}} 
                  \exp\{-\sqrt{\Delta\varepsilon_{0}} r\}
\end{equation}
and the exciton binding energy is
\begin{equation}
  \label{eq:20}
  \varepsilon_{0} =  e^{4} \Delta
  \text{.}
\end{equation}
Introducing dimensionless variables $\tilde{\varepsilon}$ and
$\tilde{\varepsilon}_{0} = \varepsilon_{0}/(2\Delta)$, we arrive at
\begin{equation}
  \label{eq:21}
  \tilde{\chi}(\tilde{\varepsilon},\qq) \approx \frac{q^{2}}{2\pi\Delta} 
                               \frac{\tilde{\varepsilon}_{0}}%
                                    {\tilde{\varepsilon}_{0}-\tilde{\varepsilon}}
\end{equation}
for $\tilde{\varepsilon}$ close to the excitonic peak of the $1s$ ground state. 

We now want to compare the position and the oscillator strength
of the excitonic peak  of the $1s$ ground state with the results
obtained from ${\fxcEx}^{(1)}$. For this we first expand $\chiQP$ of
Eq.~\eqref{eq:chiQP} for small $\tilde{\varepsilon}$ 
\begin{equation}
  \label{eq:22}
 \chiQP(\tilde{\varepsilon},\qq) \approx \frac{q^{2}}{16\pi\Delta}
             \left( \ln\left(\frac{2}{\tilde{\varepsilon}}\right) -1\right)   
 \text{.}
\end{equation}
\begin{figure}
  \includegraphics[width=\linewidth]{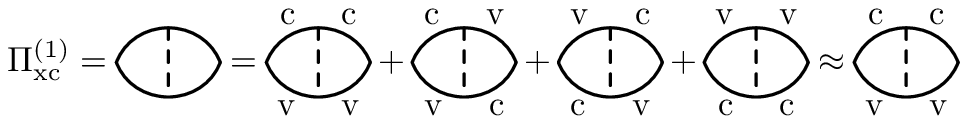}
  \caption{Approximation for $\Pixc^{(1)}$ in the limit of small
    $\tilde{\varepsilon}$.} 
  \label{fig:pixc-approx}
\end{figure}
In order to obtain ${\fxcEx}^{(1)}$ we also need an expression for
$\Pixc^{(1)}$ in the limit of small $\tilde{\varepsilon}$.  In
diagrammatic form $\Pixc^{(1)}$ is given by a single graph as displayed
in Fig.~\ref{fig:pixc-approx}. There are four possibilities to
distribute the conduction- and valence-band states in the two
particle--hole propagators. The particle--hole propagators with a
conduction-band state in the upper line and a valence-band state in the
lower line diverge at $\tilde{\omega}=1$. The particle--hole propagators
with a valence-band state in the upper line and a conduction-band state
in the lower line diverge at $\tilde{\omega}=-1$. Therefore, in the
limit where the energy goes to the band gap (or equivalently
$\tilde{\varepsilon}$ is small), the main contribution to $\Pixc^{(1)}$
comes from the diagram with both particle--hole propagators of the upper
line belonging to the conduction-band state and the lower line to the 
valence-band state. Neglecting the other contributions is our first
approximation in calculating $\Pixc^{(1)}$. The other approximation is using
Eqs.~\eqref{eq:16} and~\eqref{eq:17}. With these approximations we
obtain
\begin{equation}
  \label{eq:23}
  \Pixc^{(1)}(\tilde{\varepsilon},\qq) \approx  \frac{|\gamma(\qq,0)|^{2}}%
                            {\left(2\Delta\right)^{2}} 
     \sum_{\kk,\kk'}
     \frac{V(\kk-\kk')}%
          {\left(\frac{1}{2}\left(\frac{k}{\Delta}\right)^{2}
              +\tilde{\varepsilon}\right)
           \left(\frac{1}{2}\left(\frac{k'}{\Delta}\right)^{2}
              +\tilde{\varepsilon}\right)}
  \text{,}
\end{equation}
where $V(\kk-\kk')=2\pi e^{2}/|\kk-\kk'|$ is the Fourier
representation of 
$V(\rr-\rr')$ from Eq.~\eqref{eq:15}. The easiest way to solve this
double integral is to look at it in the real space, instead of the
Fourier space, where it becomes a single integral. The 2D Fourier
transformation of the particle--hole propagator $\left(\frac{1}{2}
  (k/\Delta)^{2} +\tilde{\varepsilon}\right)^{-1}$
gives the modified Bessel function of the second kind $K_{0}$, so that
we can write
\begin{equation}
  \label{eq:24}
  \Pixc^{(1)}(\tilde{\varepsilon},\qq) \approx
             \frac{e^{2}q^{2}}{8\pi\Delta\sqrt{2\tilde{\varepsilon}}}
             \int_{0}^{\infty} d\rho K_{0}^{2}(\rho)
  \text{.}
\end{equation}
The integral over $K_{0}^{2}$ computes to $(\pi/2)^{2}$, and we arrive
at
\begin{equation}
  \label{eq:25}
  \Pixc^{(1)}(\tilde{\varepsilon},\qq) \approx 
               \frac{q^{2}}{16\pi\Delta} \frac{\pi^{2}}{2} 
               \sqrt{\frac{\tilde{\varepsilon}_{0}}{\tilde{\varepsilon}}}
  \text{.}
\end{equation}
Now we have all ingredients to build
${\fxcEx}^{(1)}=\chiQP^{-1}\Pixc^{(1)}\chiQP^{-1}$ and insert it in
Eq.~\eqref{eq:dyson-fxcEx}. This gives the following
approximation for the response function based on ${\fxcEx}^{(1)}$:
\begin{align}
  \tilde{\chi}^{f^{(1)}}(\tilde{\varepsilon},\qq) &=
                      \frac{\chiQP(\tilde{\varepsilon},\qq)}%
                           {1-\Pixc^{(1)}(\tilde{\varepsilon},\qq)
                              \chiQP^{-1}(\tilde{\varepsilon},\qq)}
                               \nonumber\\
  \label{eq:chitilde-1-approx}
   &= \frac{q^{2}}{16\pi\Delta}
      \frac{\left( \ln\left(\frac{2}{\tilde{\varepsilon}}\right) -1\right)}%
           {1-\left( \ln\left(\frac{2}{\tilde{\varepsilon}}\right) 
            -1\right)^{-1} \frac{\pi^{2}}{2} 
               \sqrt{\frac{\tilde{\varepsilon}_{0}}{\tilde{\varepsilon}}}}
  \text{.}
\end{align}
Note that this equation is valid only for small
$\tilde{\varepsilon}$, since we derived $\chiQP$ and $\Pixc^{(1)}$ only
for small values of $\tilde{\varepsilon}$.

From Eq.~\eqref{eq:chitilde-1-approx} we clearly see that
$\tilde{\chi}^{f^{(1)}}$ contains an additional pole where the
denominator vanishes. This is the excitonic peak in this
approximation and its energy $\tilde{\varepsilon}_{0}'$ is defined by
\begin{equation}
  \label{eq:26}
  1-\left( \ln\left(\frac{2}{\tilde{\varepsilon}_{0}'}\right) 
            -1\right)^{-1} \frac{\pi^{2}}{2} 
               \sqrt{\frac{\tilde{\varepsilon}_{0}}{\tilde{\varepsilon}_{0}'}}
  =0
  \text{.}
\end{equation}
To evaluate the approximation used let us reformulate
Eq.~\eqref{eq:26} as
\begin{equation}
  \label{eq:33}
  \frac{e^{4}}{2} =
  \tilde{\varepsilon}_{0} = 2 \frac{\tilde{\varepsilon}_{0}}%
                                   {\tilde{\varepsilon}_{0}'} 
                          \exp \left(-\left(\frac{\pi^{2}}{2}
                           \sqrt{\frac{\tilde{\varepsilon}_{0}}%
                                {\tilde{\varepsilon}_{0}'}} +1\right)\right)
  \text{.}
\end{equation}
We see that the approximate binding energy $\tilde{\varepsilon}_{0}'$
and the exact binding energy $\tilde{\varepsilon}_{0}$ are identical for 
\begin{equation}
  \label{eq:27}
  \frac{e^{4}}{2} =
  \tilde{\varepsilon}_{0} = 2 \exp \left(-\left(
                                   \frac{\pi^{2}}{2}+1\right)\right)
  \approx \frac{1}{189}
  \text{.}
\end{equation}
In other words, for the interaction strength which corresponds to this 
quite realistic exciton binding energy, the approximation using
${\fxcEx}^{(1)}$ gives the exact binding energy. For stronger
and \emph{weaker} interaction strength there is some error. To be more
precise, the error is below 10\,\% for $\tilde{\varepsilon}_{0}$ between
$1/165$ and $1/222$. It is below 20\,\% for $\tilde{\varepsilon}_{0}$
between $1/148$ and $1/271$. Note that in the range of energies where
the approximation gives good results, $\tilde{\varepsilon}$ is indeed
small, so that our approximation is consistent.
It is interesting to note that for the Coulomb interaction---in
contrast to a short-range interaction---one does not get the correct
exciton binding energy from ${\fxcEx}^{(1)}$ in the limit of the
interaction strength going to zero. Note also that while the exact
solution of the 2D hydrogen problem gives rise to an infinite series of
excitonic peaks in the exact response function, we obtain only \emph{one}
excitonic peak in the approximate approach. This problem was already
touched in the introduction. Although ${\fxcEx}^{(1)}$ is frequency
dependent, it does not contain the rapid oscillations needed to describe
the whole series of excitonic states.

To compare the approximate response function of
Eq.~\eqref{eq:chitilde-1-approx} to the exact one from Eq.~\eqref{eq:21}
and determine the residual it is best to expand
$\tilde{\chi}^{f^{(1)}}$ around $\tilde{\varepsilon}_{0}'$. This
can be done by calculating the first-order Taylor expansion of the
denominator in Eq.~\eqref{eq:chitilde-1-approx}:
\begin{equation}
  \label{eq:28}
  \tilde{\chi}^{f^{(1)}}(\tilde{\varepsilon},\qq) \approx
     \frac{q^{2}}{16\pi\Delta}
     \frac{\frac{\pi^{2}}{2} 
             \sqrt{\frac{\tilde{\varepsilon}_{0}}{\tilde{\varepsilon}_{0}'}}}%
          {\frac{1}{2} - \frac{2}{\pi^{2}} 
             \sqrt{\frac{\tilde{\varepsilon}_{0}'}{\tilde{\varepsilon}_{0}}}}
     \frac{\tilde{\varepsilon}_{0}'}%
          {\tilde{\varepsilon}-\tilde{\varepsilon}_{0}'}
  \text{.}
\end{equation}
In the case where the exciton energy is exactly reproduced, \emph{i.e.}
for $\tilde{\varepsilon}_{0}' = \tilde{\varepsilon}_{0}$, the second
fraction in Eq.~\eqref{eq:28} is about $16.6$. The oscillator strength
is thus too large by a factor of two inspite of the fact that the
binding energy is exact. The value of this second fraction is almost
constant for a large range of ratios
$\tilde{\varepsilon}_{0}'/\tilde{\varepsilon}_{0}$, so that mainly the
$\tilde{\varepsilon}_{0}'$ in the numerator of the third fraction in
Eq.~\eqref{eq:28} produces additional errors in the oscillator strength.
The larger oscillator strength is not too surprising, as the excitonic
oscillator strength is distributed over more peaks in the exact response
function.

Similar to the results for the short-range interaction of the previous
section, we observe that the cancellation effects are also effective for
the long-ranged Coulomb interaction. However, the general structure of
this cancellation is different and cannot be universally characterized.
Hence the accuracy of ${\fxcEx}^{(1)}$ has to be checked for every
particular system. The integral equation for $\Lambda$ together with the
replacement procedure of Fig.~\ref{fig:bse-lambda} provides an
appropriate tool for this task. 
  

\section{Conclusions}
\label{sec:conclusion}

In conclusion, we have investigated the excitonic effects on the response
function within TDDFT. We have split the xc~kernel into a
quasiparticle~$\fxcQP$ and an excitonic part~$\fxcEx$. Using a
diagrammatic expansion we derived the integral equation for the
three-point function $\Lambda$, in terms of which $\fxcEx$ can be
exactly expressed. As this integral equation is similar in structure to
the BSE, it establishes the connection between the common many-body
theory and TDDFT. The kernel of the equation for $\Lambda$ shows the possibility of
cancellation effects. If these cancellations were complete, $\fxcEx$
would be equal to its first-order approximation ${\fxcEx}^{(1)}$. This
suggests that in some situations ${\fxcEx}^{(1)}$ provides a good
substitute to $\fxcEx$. 

We have presented explicit calculations for a model two-band
semiconductor with short-ranged interaction which can be solved
analytically. Comparing the exact response function with the response
function derived from ${\fxcEx}^{(1)}$ we confirm a very good agreement
for a weak interaction, where both the position of the shallow exciton
and the Sommerfeld factor are correctly described. We were able to trace
this to the strong cancellation occuring for energies close to the band
gap (almost) independently of the interaction strength. Calculations
with the Coulomb interaction give similar results, though the
deficiencies of ${\fxcEx}^{(1)}$ are somewhat worse in this case.

These calculations represent an example that the integral
equation for $\Lambda$ can serve as a tool for evaluation of the
validity of low-order approximations to $\fxcEx$. Testing the
predictive power of this approach for other systems will be the subject
of further work.

This work was supported by the Deutsche Forschungsgemeinschaft under
Grant No. PA 516/2-3.


\end{document}